\documentclass{elsart}

\usepackage{epsfig}
\usepackage{cite}

\usepackage{amssymb}

\def\be{\begin{equation} }
\def\ee{\end{equation} }
\def\ba{\begin{eqnarray} }
\def\ea{\end{eqnarray} }
\def\ban{\begin{eqnarray*} }
\def\ean{\end{eqnarray*} }

\def\arco{\mbox{ArCO$_2$}}

\def\ExB{\mbox{\boldmath$\rm E \times B$\unboldmath}}
\def\mum{\mbox{$\mu$m}}
\def\mumrcm{\mbox{$\mu$m / $\sqrt{\mbox{cm}}$}}

\begin{document}

\begin{frontmatter}


\title{Resolution studies of cosmic-ray tracks in a TPC with
GEM readout}

\author[ref_CU]{R.~K.~Carnegie},
\author[ref_CU,ref_TR]{M.~S.~Dixit},
\author[ref_CU]{J.~Dubeau},
\author[ref_CU,ref_UVIC,ref_TR]{D.~Karlen},
\author[ref_UM]{J.-P.~Martin},
\author[ref_CU,ref_TR]{H.~Mes}
and
\author[ref_CU]{K.~Sachs\corauthref{add_KS}}
\ead{sachs@physics.carleton.ca}

\address[ref_CU]{Department of Physics, Carleton University, 
        \\ 1125 Colonel By Drive, Ottawa, ON, K1S 5B6, Canada}
\address[ref_UM]{University of Montreal, Montreal, QC, Canada}
\address[ref_UVIC]{University of Victoria, Victoria, BC, Canada}
\address[ref_TR]{TRIUMF, Vancouver, BC, Canada}
\corauth[add_KS]{Corresponding author; 
         tel.: +1-613-520-2600, ext. 1567; fax: +1-613-520-7546.}


\begin{abstract}
A large volume TPC is a leading candidate for the central tracking
detector at a future high energy linear collider. To improve the
resolution a new readout based on micro-pattern gas detectors is
being developed. Measurements of the spatial resolution of
cosmic-ray tracks in a GEM TPC are presented.
We find that the resolution suffers if the readout pads
are too wide with respect to the charge distribution at the 
readout plane due to insufficient charge sharing. For narrow pads of
$2 \times 6 \mbox{ mm}^2$ we measure a resolution of 100 \mum\ at 
short drift distances in the absence of an axial magnetic field. 
The dependence of the spatial resolution as a function of drift 
distance allows the determination of the underlying electron statistics.
Our results show that the present technique uses about half 
the statistical power available from the number of primary electrons.
The track angle effect is observed as expected.
\setlength{\unitlength}{1mm}
\begin{picture}(0,0)
\put(-20,160){\parbox{5cm}{NIM A538 (2005) 372-383 \\ physics/0402054 \\ LC-DET-2004-004}}
\end{picture}
\end{abstract}

\end{frontmatter}

\section{Introduction}
\label{sec:intro}

The time projection chamber (TPC) \cite{cit:TPC1,cit:TPC2} has been a 
mainstay of large particle detectors since its initial 
concept in the 1970's. The traditional TPC has an end cap  
detector that uses anode wires for amplification of the signal. 
When operated in an axial magnetic field, this leads 
to the so called \ExB\ effect \cite{cit:ExB} close to the wires,
which significantly degrades the resolution of the TPC.
Proposals to readout TPC signals without the use of anode wires have 
been suggested \cite{cit:padTPC1,cit:padTPC2} in the past. 
The recent development and success of micro pattern gas detectors 
(MPGD) such as the $\mu$Megas \cite{cit:uMegas} and the 
GEM \cite{cit:gem,cit:gem2} has renewed interest in this solution.

The advantages of MPGD detectors are that they require less mass 
for construction, should not have any \ExB\ effect, naturally suppress 
positive ion feedback and allow more freedom in the shape and 
orientation of the readout pads. In addition the signals are faster,
potentially improving the double track resolution in drift time.
In the case of MPGDs, the signal on 
the readout pads can be a direct electron collection signal or an induced signal. 
The advantage of direct signals is that their amplitude is larger and 
the signal is more confined, thus potentially improving the spatial
double track resolution. The disadvantage of the confined signal is 
that the pads need to be much narrower, on the order of the width of 
the ionization charge distribution, which increases the number of 
channels and thus the cost. In the case of GEMs the ionization charge 
can be spread naturally in the gaps between the GEMs and readout pads. 
It is also possible to use the induced signal 
\cite{cit:induced,cit:LCWS2000} which has a wider spread than the 
direct signal, but a reduced amplitude.

GEM amplification with pad type readout planes has been shown to give 
excellent spatial resolution for point sources, such as X-rays 
converting in a gas \cite{cit:xray2}, which is useful for 
medical applications, where the pad size can be arbitrarily small to 
give the required resolution. In the case of a large scale experiment 
using a TPC, such as the proposed TESLA detector, the pad size 
determines the number of channels and thus the cost; in that case it 
is important to make the pad size as large as possible consistent with 
the resolution required.

In earlier studies \cite{cit:LCWS2000}, using a double GEM amplification 
stage, we determined the point resolution, $s$, that can be 
achieved for X-rays converting in the gas using the direct charge 
distribution near the edge of hexagonal pads \mbox{($s \sim 70\;\mum$)} 
and the induced charge distribution near the middle of pads 
\mbox{($s \sim 80\;\mum$)}. 
A subsequent study \cite{cit:LCWS2002} with cosmic rays and 
a small TPC with an end cap detector with 5 staggered rows of 
$2.5 \times 5 \mbox{ mm}^2$ rectangular pads showed that these pads 
produced an adequate track resolution using the direct charge. 

In this paper we examine the resolution that can be achieved using 
the direct signal from a double GEM amplification stage and a 
rectangular staggered pad readout scheme. In particular we examine the 
effect of the pad width and length on the spatial resolution and 
attempt to gain a better understanding of the phenomena that affect 
the resolution. For this purpose we
measured the spatial resolution as a function of several different
quantities, including three different pad sizes 
and local position across a pad, two gases, drift distance, 
crossing angle, and signal amplitude. 

The two gases used were P10  (Ar(90):CH$_4$(10)), a fast gas with 
large  diffusion, and Ar(90):CO$_2$(10), a slow gas with relatively 
small diffusion, operated at a voltage below the peak velocity. 
The different diffusion properties allowed us to study the effect of 
pad size relative to the width of the direct charge distribution 
arriving at the pads, and to simulate, with the \arco\ mixture, 
reduced diffusion as would be present with a P10 type gas in a 
magnetic field.

\section{Experimental setup}
\label{sec:exp}

The test TPC used for these measurements is housed in a cylindrical
pressure vessel filled with P10 or \arco\ gas at atmospheric pressure.
The TPC has a maximum drift length of 15~cm and an active area of
$8 \times 8 \mbox{ cm}^2$. The drift field of 138~V/cm is established
by a series of thin window frame electrodes located between the
cathode plane at the far end and the readout end plane at the other end 
of the TPC.
A charged particle crossing the drift region will ionize the gas;
the released electrons drift to the end plane where they are amplified
and detected on a readout PCB. While drifting the charge cloud gets 
wider due to transverse diffusion, an effect that would be reduced 
in an axial magnetic field.

We use a double GEM structure for amplification with a gap
of 2.4~mm between the first and the second GEM. The voltage difference
across this transfer gap is 653~V resulting in a field of 2.7~kV/cm.
The induction gap between the second GEM and the readout board is 
5.4~mm wide with a voltage difference of 1783~V and a field of 
3.3~kV/cm. The transfer and the induction gaps were purposely large 
to diffuse the electron cloud and thus spread the signal over 
more than one readout pad.

\begin{figure}[b]
\centerline{\mbox{\epsfxsize=10cm \epsffile{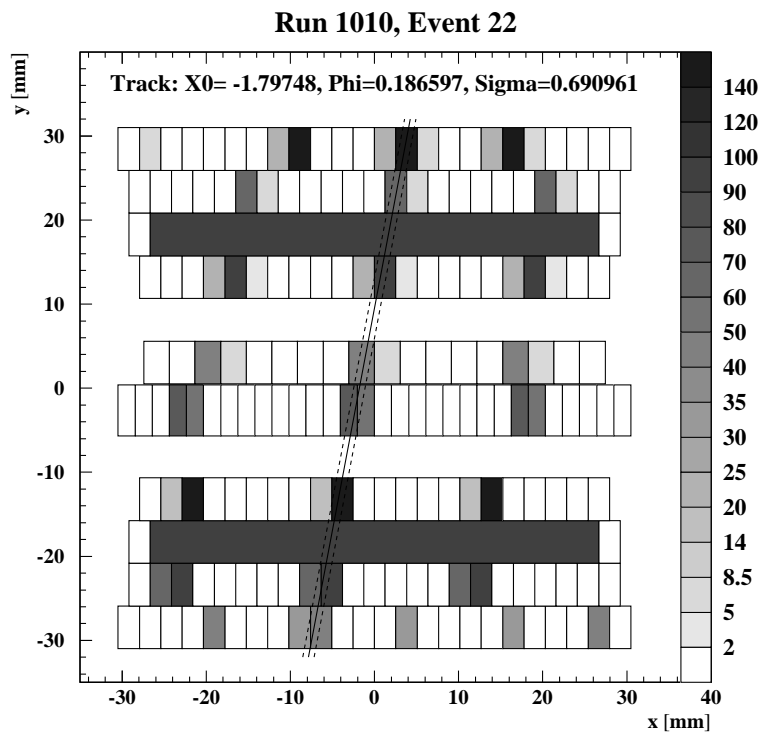}}}
\caption[]{\label{fig:event}
Event display with a reconstructed track in \arco\ gas. The shade of the
pad corresponds the the reconstructed signal amplitude. The lower
threshold is 2 and one hit per row of more than 8.5 is required for
the track fit. In general there is a 3-fold multiplexing for
the outer and test rows. The rows are numbered from bottom to top.} 
\end{figure}

The GEMs were made from 50 \mum\ thick kapton foil coated with copper 
on both sides. The holes with a diameter of $\sim$90 \mum\ at
the surface are arranged in a 
hexagonal pattern with a spacing of $d_{\rm hex} = 140$ \mum.
The voltage across the GEMs is 357~V each. 
Gain measurements were made for 5.9 keV $^{55}$Fe  x-ray conversion
electrons in the gas and used standard pulser calibration technique. 
The effective gains for \arco\ and P10 were about 6700 and 4600 respectively.

The readout-pad layout shown in Figure \ref{fig:event} consists 
of 192 pads which are reduced to 64 readout channels via 
multiplexing. The pads in the outer rows (1,2,4,7,9,10) are
2.54 mm $\times$ 5.08 mm large; in the test row 5 the pads are
2.032 mm $\times$ 6.096 mm large and in row  6 they are
3.048 mm $\times$ 5.080 mm large.
Rows 3 and 8 consist of wide pads used for filtering. The outer
pads in rows 1--4 and 7-10 are multiplexed
to give one veto channel on the left and right side, respectively.

We use a right-handed coordinate system with the $x$-coordinate
horizontal and the $y$-coordinate pointing upwards; the 
$z$-coordinate corresponds to the drift distance with $z=0$ at the
first GEM. The azimuthal angle $\phi$ and the polar angle $\theta$
are measured with respect to the $y$-axis.

The signals are read out via ALEPH TPC wire 
preamplifiers \cite{cit:preamp} and
64 channels of 200~MHz, 8~bit FADCs custom made at the University of Montreal.
For data acquisition we use the MIDAS \cite{cit:MIDAS} system.

A three layer scintillation counter telescope is used to trigger on 
cosmic-ray tracks.
One scintillator counter is placed above the TPC and two below, 
separated by a 10 cm thick layer of lead. The $\sim$19 cm width of 
the counters and the distance of $\sim$40 cm between the two outer 
counters defines the acceptance coverage in z.  

\section{Theory}
\label{sec:theory}

The observed width\footnote{Throughout this paper the width of a 
distribution refers to its standard deviation.}
of the track, $\sigma_{\rm track}$, which is the standard deviation 
of the charge cloud perpendicular to the track, is determined by
two components, the spread associated with the readout system $\sigma_0$ and
the transverse diffusion depending on the drift distance $z$:
\ba
\label{eq:trackwidth}
\sigma_{\rm track}^2 & = & \sigma_0^2 + C_{\rm D}^2 \; z \\
\nonumber
\sigma_0^2 & = &\sigma_{\rm hex}^2 + \sigma_{\rm intern}^2 
                                   + \sigma_{\rm other}^2 \; ,
\ea
where $C_{\rm D} = \sqrt{2 D_{\rm t} / \nu}$ is the 1-dimensional 
diffusion coefficient given by the transverse diffusion constant 
$D_{\rm t}$ and the drift velocity $\nu$. In a 
magnetic field $D_{\rm t}(B) = D_{\rm t}(0) / (1+\omega^2 \; \tau^2)$,
thus resulting in a reduced transverse diffusion.
The contribution $\sigma_0$ is composed of several parts.
The first term originates from the hexagonal pattern structure
of the GEM depending on the hole distance $d_{\rm hex}$. For our
geometry $\sigma_{\rm hex}$ is estimated to be $\sim$ 50 \mum .
The second term, $\sigma_{\rm intern}$, results from diffusion 
between the GEMs and the readout pads. For the present setup
\mbox{$\sigma_{\rm intern}\simeq  318\;\mum$} for \arco\ and 
$\simeq 460\;\mum $ for P10. Other factors denoted by 
$\sigma_{\rm other}$ also contribute.

The standard deviation $\sigma_x$ of the charge cloud distribution
across a row of pads also 
includes the crossing angle effect $\sigma_\phi$ depending on
the track angle $\phi$ and is given by: 
\ba
\label{eq:sigmax}
\sigma_x^2 & = & \sigma_0^2 + \sigma_{\rm D}^2 + \sigma_\phi^2 \\
\nonumber
\sigma_{\rm D} & = & C_{\rm D} \; \sqrt{z} / \cos{\phi} \\
\nonumber
\sigma_\phi & = & L/\sqrt{12} \; \tan{\phi} \; .
\ea

The factor $1 / \cos{\phi}$ in the transverse diffusion  
term $\sigma_{\rm D}$ reflects the projection of the charge 
distribution onto the x-axis. 
The crossing angle effect comes from the spread of $x(y)$ for
a track with an angle $\phi$. Projected onto the $x$ axis
this leads to a rectangular function of total width 
$L\tan{\phi}$, where $L$ is the length of the pad. 
The standard deviation for such a rectangular uniform 
distribution is given by $\sigma_\phi$.

The observed x-resolution $s_x$ reflects the precision with which  
the mean of the charge distribution can be determined
and hence has additional factors from statistics:
\ba
\label{eq:xreso}
s_x^2 & = & s_0^2 + s_{\rm D}^2+ s_\phi^2  \\
\nonumber
s_0^2 & = & s_{\rm hex}^2 + s_{\rm intern}^2 + s_{\rm other}^2\\
\nonumber
s_{\rm D} & = & \sigma_{\rm D} / \sqrt{N^{\rm eff}_{\rm t}}  \\
\nonumber
s_\phi & = & \sigma_\phi / \sqrt{N^{\rm eff}_{\rm cl}}  \; .
\ea

Most contributions depend on the number of electrons $n_{\rm t}$
produced by the ionizing particle. Some of these electrons stem from 
secondary ionization. They are therefore correlated forming $n_{\rm cl}$
clusters. The number of electrons and clusters created across a
row of pads is \mbox{$N_{\rm t} = n_{\rm t} \cdot L / \cos{\phi}$} and
\mbox{$N_{\rm cl} = n_{\rm cl} \cdot L / \cos{\phi}$}, respectively.
For example for Argon \mbox{$n_{\rm cl} = 24.3 / \mbox{cm}$} and
\mbox{$n_{\rm t} = 94 / \mbox{cm}$} \cite{cit:kkk}.

All components of $s_0$ depend on the signal amplitude.
The contribution from the GEM structure $s_{\rm hex}$ is minor:
the track width due to diffusion is large enough to cover a 
sufficient number of GEM holes. The effect from internal diffusion 
$s_{\rm intern}$ is strongly reduced due to the high gain.
Contributions from electronic noise, calibration errors and
limitations due to insufficient charge sharing between the pads
are included in $s_{\rm other}$.

The effect from transverse diffusion depends on the effective number
of electrons $N_{\rm t}^{\rm eff} = R \cdot N_{\rm t}$, which includes 
a reduction factor $R$.
The crossing angle effect depends on the effective number of clusters 
$N^{\rm eff}_{\rm cl} =  (N_{\rm cl})^\epsilon$.
Following the notation of \cite{cit:blum} the number of clusters
is reduced by the exponent $\epsilon$.

\begin{figure}[b]
\centerline{\mbox{\epsfxsize=10cm \epsffile{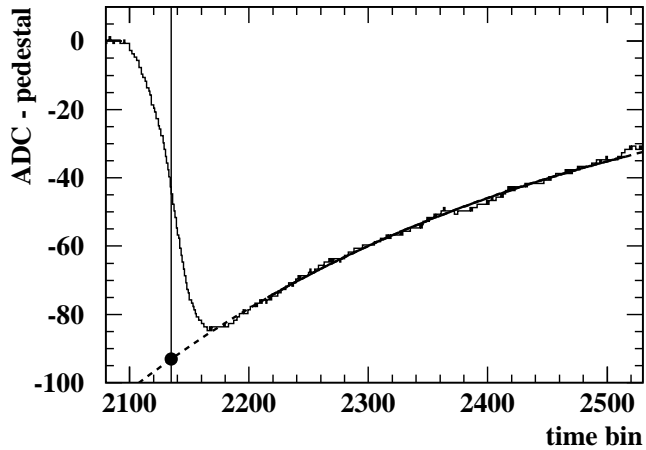}}}
\caption[]{\label{fig:pulse}
Determination of time and amplitude of a pulse, see text. 
The dot indicates the reconstructed $T0$ and amplitude.}
\end{figure}

\section{Reconstruction}
\label{sec:rec}

\begin{figure}[b]
\centerline{\mbox{\epsfxsize=10cm \epsffile{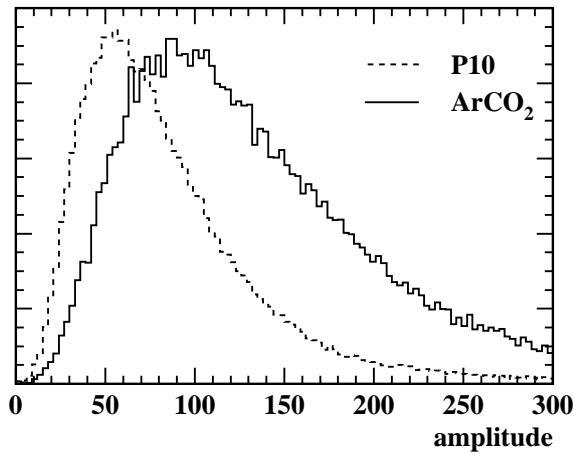}}}
\caption[]{\label{fig:amplitude}
Sum of pad signal amplitudes in row 5 for both gases.
The distribution is proportional to the energy loss, the difference
between \arco\ and P10 is due to different gain.
The RMS of the noise on a single pad is $\sim$0.5 ADC counts with
typically 2 to 3 pads contributing to the amplitude of the row.
The cutoff for a pad signal was 2 ADC counts.}
\end{figure}

The analysis package is based on Fortran f95 code \cite{cit:F}.
In a calibration run pedestals and pulse fall times $t_{\rm fall}$
as well as the relative gain are determined for each readout channel. 
The time $T0$ and the amplitude of the signals are
determined from the pedestal subtracted ADC pulse as shown in 
Figure~\ref{fig:pulse}. 
The time $T_{\rm peak}$ is determined as the time bin with 
minimum ADC count. In the region $[T_{\rm peak}+50 ; T_{\rm peak}+350]$
an exponential 
$A(t) = A_{\rm peak} \ast \exp{-(t-T_{\rm peak})/t_{\rm fall}}$
is fit to the ADC spectrum to determine the amplitude $A_{\rm peak}$
at $T_{\rm peak}$. The time $T0$ is determined via a linear fit to
25 time bins at the rising edge as $\mbox{ADC}(T0) = A_{\rm peak} / 2$ 
and the signal amplitude is $-A(T0)$. The amplitudes are corrected
for the relative gain of each channel. The RMS of the correction
coefficients is 5\%.
Only signals with an amplitude of more than 2 ADC counts are 
recognized as pad hits and taken into account.
Events are rejected if a veto channel has an amplitude of more than 
8.5 ADC counts.
The $T0$ of a row is determined as the amplitude weighted mean
of the times of the hits in this row.
The sum of reconstructed amplitudes in row 5 is shown in Figure 
\ref{fig:amplitude}. 

The track fit is performed similar to \cite{cit:deanfit}.
In the upper and lower two rows (1,2 and 9,10) start points are
determined from a centroid calculation of the largest amplitude
channel and its neighbor pads. These two points are connected
by a line to form the seed track. Because of the multiplexing
several seed tracks are found and the track with the most rows 
having a related hit with an amplitude of more than 8.5 
is chosen. In general this choice is unique. There must be at least
six rows with hits out of the eight outer and test rows.
Other events are rejected from the analysis.

The track projection in the x-y plane can be described with three 
parameters: the x-position
at y=0, $x_0$, the track angle, $\phi$, and the width of the charge
cloud, $\sigma_{\rm track}$. 
The track parameters are determined from a maximum likelihood fit where a
uniform line of charge with a Gaussian profile is assumed. This
idealized distribution is integrated over the pad areas and normalized
across a row to obtain the expected charge probabilities. From these 
and the observed signal amplitudes a likelihood function is determined,
which includes a uniform noise probability of 0.2\%. The noise level is 
determined from the data; a variation between 0.1 and 0.5\% has only a 
small effect on the fitted track parameters.

The drift distance at y=0, $z_0$, and the angle
$\theta$ are determined from a straight line fit to the $T0$ of each
row as a function of y. All eight rows are used to determine global
distributions of track angles $\phi$ and $\theta$ as well as
$x_0$ and $z_0$.
The drift velocities as determined from the data are 55$\pm$4 \mum /ns 
for P10 and \mbox{8.3$\pm$0.3 \mum /ns} for \arco .
The result for P10 is in good agreement with the prediction from
MAGBOLTZ given in Table \ref{tab:magboltz}, 
while the measured velocity for \arco\ is smaller than expected. 
This might be due to a limited time window recorded or
because of gas impurities. In \arco\ we lose about 10\% of the 
electrons over the 
full drift distance of 15 cm due to attachment. This is an indication
of impurities in the gas which may affect gas properties.
No such effect is observed with P10.

\begin{table}[t]
\caption[]{\label{tab:magboltz}
MAGBOLTZ (version 5.1) predictions for a drift field of 140 V/cm.}
\begin{center}
\begin{tabular}{l||c|c|c}
\hline
    & drift & \multicolumn{2}{c}{diffusion} \\
gas & velocity & transverse & longitudinal \\
    & (\mum /ns) & (\mumrcm ) & (\mumrcm ) \\
\hline\hline
P10 & 55 & 564 & 374\\
\arco & 8.9 & 229 & 241 \\
\hline
\end{tabular}
\end{center}
\end{table}

\begin{figure}[t]
\centerline{\mbox{\epsfxsize\textwidth\epsffile{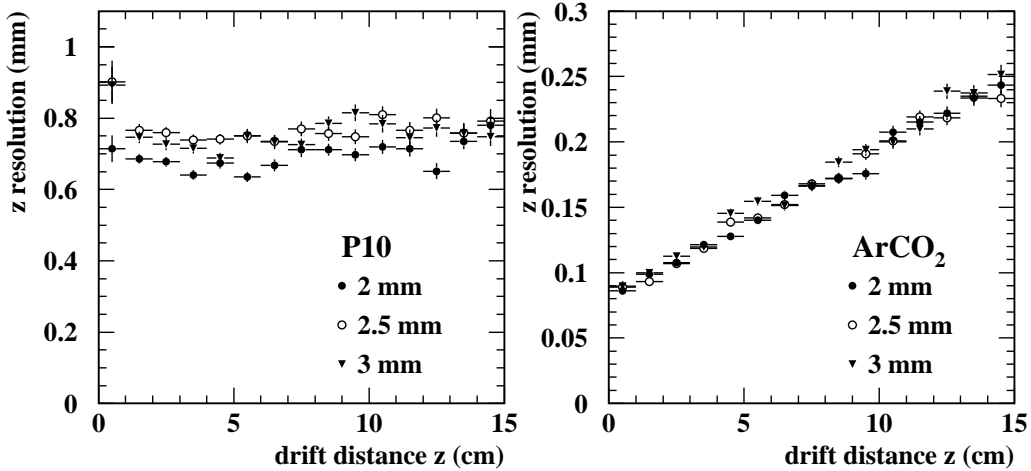}}}
\caption{\label{fig:zreso} 
Resolution of drift distance for both gases as a function
of drift distance for small track angle $|\phi| < 5^\circ$. }
\end{figure}

\section{Analysis Results}
\label{sec:results}

The resolution for the drift distance $z$ is shown in Figure \ref{fig:zreso}
as a function of drift distance. The intrinsic time resolution
is about 13~ns for P10 and 9~ns for \arco . It is worse for
P10 since the average signal amplitude is smaller.
While the $z$ resolution for P10 is completely dominated by the intrinsic
time resolution, the effect of longitudinal diffusion is visible
for \arco\ because of the much smaller drift velocity.
The observed dependence is linear and not quadratic as expected:
$s_z / \mum = 80 + 14 * z / \mbox{cm}$.
It does not depend on the readout pad size.
 
\begin{figure}[t]
\centerline{\mbox{\epsfxsize\textwidth\epsffile{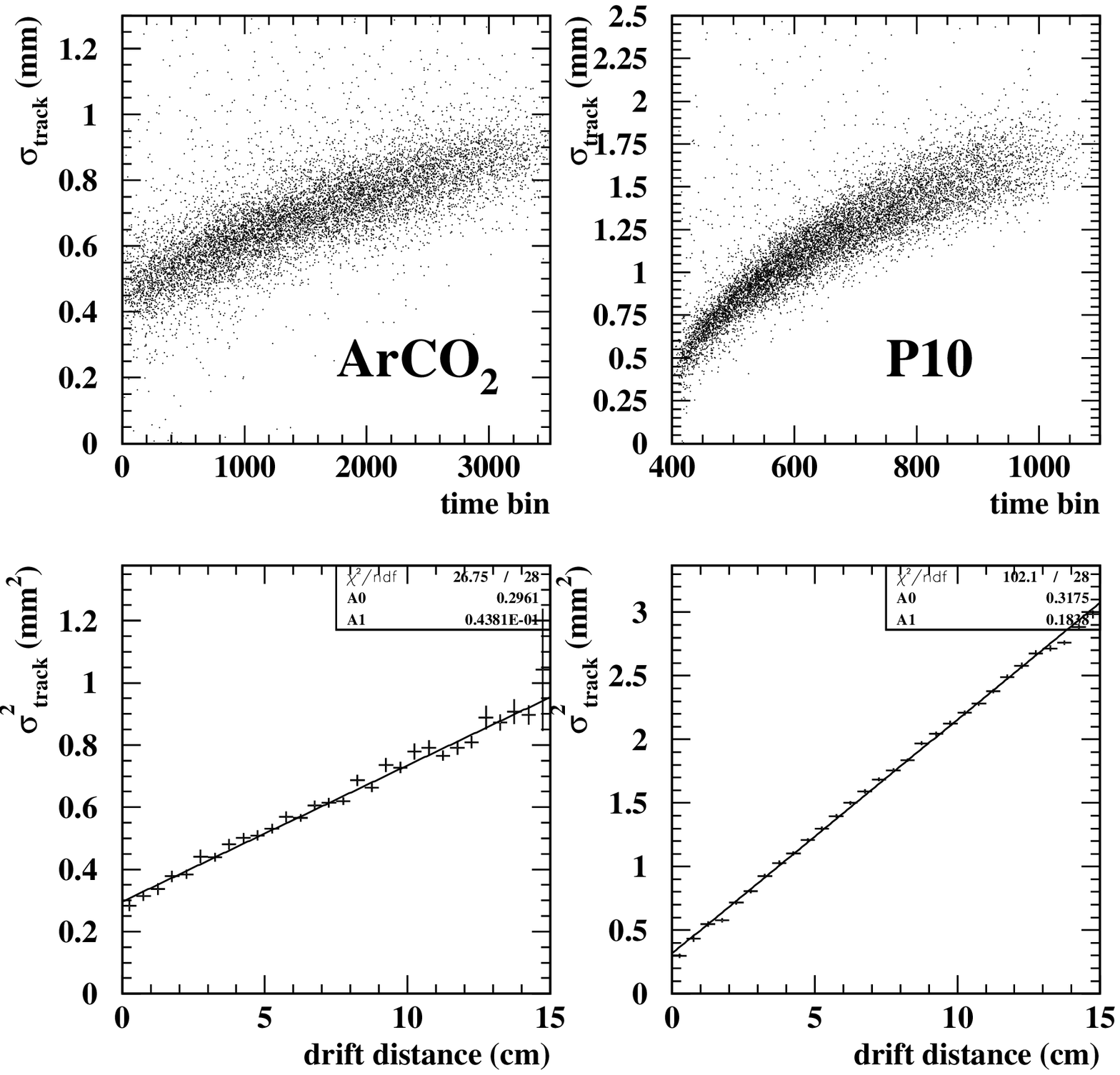}}}
\caption[]{\label{fig:twidth}
Fitted track width $\sigma_{\rm track}$ as a function of drift distance $z$. 
The upper plots show $\sigma_{\rm track}$ versus $z$ for all events,
the lower plots show the average $\sigma_{\rm track}^2$ versus $z$. }
\end{figure}

The width of the charge cloud  $\sigma_{\rm track}$ is shown in Figure
\ref{fig:twidth} as a function of the drift distance. The mean
transverse diffusion coefficient can be determined from a linear fit to
$\sigma_{\rm track}^2(z)$ (Equation \ref{eq:trackwidth}). We obtain 
\ba
\label{eq:CDp10}
C_{\rm D} & = & 0.429 \mbox{ mm} / \sqrt{\mbox{cm}} 
\hspace{10mm}\mbox{for P10 and} \\
\label{eq:CDarco}
C_{\rm D} & = & 0.209 \mbox{ mm} / \sqrt{\mbox{cm}} 
\hspace{10mm}\mbox{for \arco} \; ,
\ea
with negligible statistical errors. The result for P10 is 
smaller than the expectation from MAGBOLTZ given in Table 
\ref{tab:magboltz} while the result for \arco\ is in reasonable
agreement. The width of the charge cloud at $z=0$ is 
determined to be $0.563 \pm 0.006 \mbox{ mm}$ for P10 and
$0.544 \pm 0.006 \mbox{ mm}$ for \arco .
Only a part of this measured amount is expected from 
$\sigma_{\rm intern}$. Since $\sigma_{\rm hex}$ is negligible 
a contribution of several 100 \mum\ remains unexplained for both gases
and must be assigned to other factors in $\sigma_{\rm other}$.

The remainder of this paper concentrates on the study of the $x$ resolution
dependence on track angle, transverse diffusion and amplitude. 
In this analysis the track parameters are not determined from
reconstructed hit-positions in each row but from a fit to the 
charge distribution of the full track. Therefore the concept of
the point position in a row is not a priori given. The x-position
in a row, $x_{\rm row}$ is determined from a track fit to the
charge distribution in this row
only, keeping all track parameters fixed apart from $x$.
The x-resolution $s_x$ is obtained from the width~$\sigma$ of a 
Gaussian fit to the distribution of the residuals 
$\delta = x_{\rm row} - x_{\rm track}$; 
$x_{\rm track} = x_0 + \tan{\phi}*y_{\rm row}$. 
If these residuals are derived from a track fit including the test row
the obtained spread $\sigma^{\rm in}$ will be systematically too
small. On the other hand the spread $\sigma^{\rm ex}$ obtained 
from a track fit excluding the test row will be too large.
As shown in appendix \ref{sec:rowfit} the geometric mean of these
two results $s_x = \sqrt{\sigma^{\rm in} * \sigma^{\rm ex}}$
gives the correct estimate for the point resolution.
For these studies the charge width $\sigma_{\rm track}$ is fixed to the mean 
observed track width as a function of the drift distance.

\subsection{X resolution depending on pad width}

\begin{table}[t]
\caption[]{\label{tab:drift}
Mean track width and RMS for three regions of drift distance and
two gases.}
\begin{center}
\begin{tabular}{c|cc}
\hline
 & \multicolumn{2}{c}{track width (mm)} \\
drift distance & \arco & P10 \\
\hline
0 --  3 cm & 0.53$\pm$0.14 & 0.72$\pm$0.24 \\
3 --  8 cm & 0.67$\pm$0.13 & 1.12$\pm$0.21 \\
8 -- 15 cm & 0.81$\pm$0.11 & 1.52$\pm$0.21 \\
\hline
\end{tabular}
\end{center}
\end{table}

\begin{figure}[b]
\centerline{\mbox{\epsfxsize=10cm \epsffile{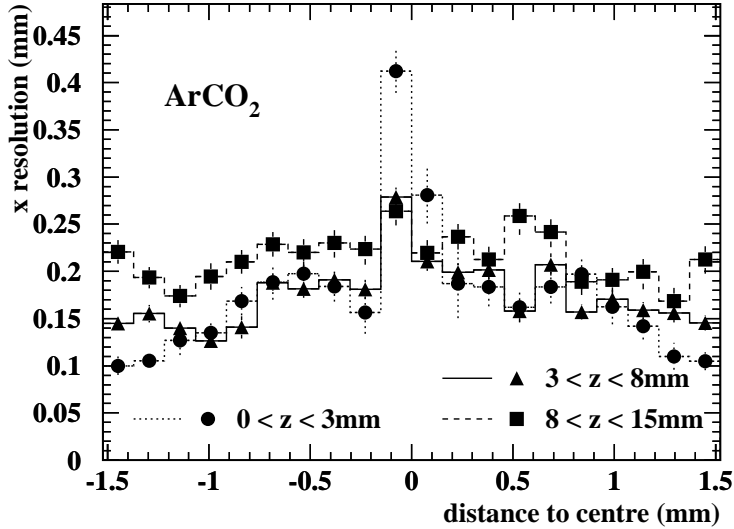}}}
\caption[]{\label{fig:edge}
X resolution as a function of distance to the centre of the pad
for \arco\ and three regions of drift distance. 
The points are results from 3 mm wide pads with $|\phi|<5^\circ$. }
\end{figure}

First we investigate the dependence of the resolution on the 
width of the pad. To eliminate other effects only tracks
with $|\phi|<5^\circ$ are used for this study.
Figure \ref{fig:edge} shows the $x$ resolution in 3 mm wide pads
as a function of the distance between the reconstructed position and 
the centre of the pad for \arco . To obtain samples with different 
diffusion, i.e. size of the charge cloud, three regions of 
drift distance are considered as given in Table \ref{tab:drift}. 
For short drift distances, hence small charge-cloud size, 
the resolution gets significantly worse in the centre of the pad.
This is because an increased fraction of signals is collected 
only on one pad and charge sharing is less effective
for the determination of the position of the track in this row. 
This effect leads also to a non-uniform distribution of the
measured $x_{\rm row}$ positions, where more hits are reconstructed 
in the center of a pad if the pad is too wide. 
This study is repeated for the 2~mm and 2.5~mm (row~4) wide pads,
which shows that the effect sets in if the pad is wider than about 
three times the width of the charge cloud and becomes prominent 
for a pad width larger than four times the charge width.
However, this effect depends also on the amplitude; 
signals with large amplitude tend to have more charge sharing.  
All measurements are made without magnetic field. However,
the results can be reinterpreted for a given width of the charge cloud.
This study indicates that 3~mm wide pads are too wide for a charge
cloud with a width of less than about 1~mm, which is the case for 
\arco\ and P10 at small drift distances. To avoid this 
effect, the following analyses are restricted to the 2~mm wide pads.

\subsection{X resolution depending on drift distance $z$}

\begin{figure}[b]
\vspace*{6mm}
\centerline{\mbox{\epsfxsize=10cm \epsffile{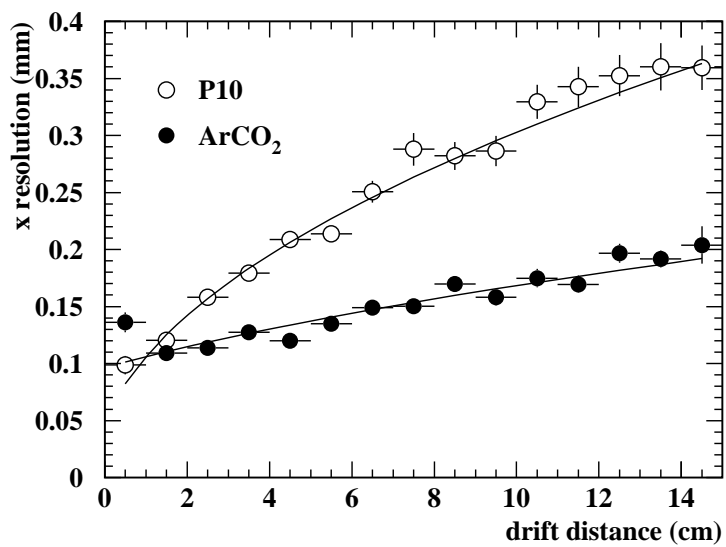}}}
\caption[]{\label{fig:drift}
Resolution in $x$, $s_x$, as a function of drift distance for 
2 mm wide pads, tracks with $|\phi|<5^\circ$, both gases.
 }
\vspace*{5mm}
\end{figure}

The resolution deteriorates with increasing drift distance because 
of transverse diffusion. This is studied using tracks with small
angle, $|\phi|< 5^\circ$, to suppress the track angle effect. 
As can be seen in Figure \ref{fig:drift}
the effect is less pronounced in \arco\ because of the very small 
diffusion. The function
\be
s_{\rm x} = \sqrt{ s^2 +  \frac{C_{\rm D}^2}{N_{\rm t}^{\rm eff}} z}
\ee
can be fit to the data. Using $C_{\rm D}$ as given in
Equations \ref{eq:CDp10} and \ref{eq:CDarco}
we obtain $N_{\rm t}^{\rm eff} = $ 19$\pm$7 for \arco\ 
and 20.6$\pm$0.7 for P10. These numbers are smaller than 
the total expected number of electrons $N_{\rm t}$. However, the wide
range of amplitudes makes the interpretation difficult.
The expected $N_{\rm t}$ as given in \cite{cit:kkk} relates
to the mean amplitude observed in a row, while the resolution 
is proportional to $1/\sqrt{N_{\rm t}}$, which results in a bias. For 
the full range the mean amplitude $\langle A \rangle$ is larger than 
$1/(\;\langle 1/\sqrt{A} \rangle\;)^2$ by 30\% for both gases.

\begin{table}[bp]
\caption[]{\label{tab:transpa}
Fit result for resolution as a function of drift distance. 
The number of electrons 
$N_{\rm t}$ for tracks with small angle $|\phi|$ is
scaled with mean amplitude; for the full range $n_{\rm t}$ is
taken from \cite{cit:kkk}.
The errors are statistical only.}
\begin{center}
\begin{tabular}{cc|c|r@{$\pm$}l|r@{$\pm$}lr@{$\pm$}l}
\hline
\multicolumn{2}{c|}{Amplitude}& &
\multicolumn{2}{c|}{$C_{\rm D}/\sqrt{N_{\rm t}^{\rm eff}}$} &
\multicolumn{4}{c}{derived}\\
range & mean & $N_{\rm t}$ &
\multicolumn{2}{c|}{(\mumrcm ) } & 
\multicolumn{2}{c}{$N_{\rm t}^{\rm eff}$} &
\multicolumn{2}{c}{$R   = \frac{N_{\rm t}^{\rm eff}}{N_{\rm t}}$ (\%)}\\
\hline
\multicolumn{9}{l}{P10} \\
all              & 93  &  55 &
 94.6&1.6 &  20.6&0.7  & 
 \multicolumn{2}{c}{}\\ 
  $0 < A < 60$   & 42  &  25 &
122.3&2.6 &  12.3&0.5  & 50.0&2.1  \\
 $60 < A < 100$  & 78  &  46 &
 93.9&2.0 &  20.9&0.9  & 45.3&1.9  \\
$100 < A < 1000$ & 175 & 103 &
 70.0&3.1 &  37.6&3.3  & 36.3&3.2  \\
\hline
\multicolumn{9}{l}{\arco} \\
all              & 150 &  57 &
 47.8&9.3 & 19&7 & 
 \multicolumn{2}{c}{}\\ 
~~$0 < A < 100$~ & 69  &  26 &
 55.6&3.0 & 14&2 &  54&6 \\
$100 < A < 170$~ & 131 &  50 &
 40.2&2.8 & 27&4 &  55&8 \\
$170 < A < 1000$ & 279 & 106 &
 23.2&3.9 & 82&27 & 77&26  \\
\hline
\end{tabular}
\vspace*{6mm}
\end{center}
\end{table}

\begin{figure}[b]
\centerline{\mbox{\epsfxsize=10cm \epsffile{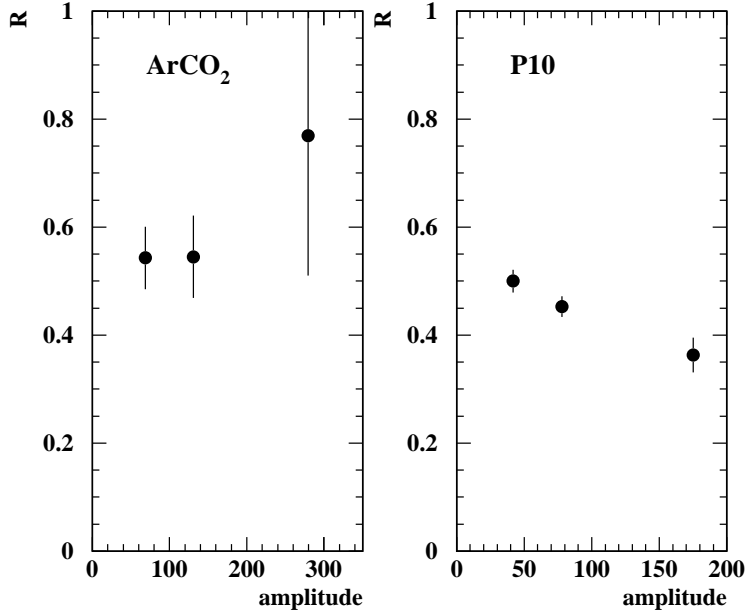}}}
\caption[]{\label{fig:transpa}
Ratio $R   = N_{\rm t}^{\rm eff} /N_{\rm t}$ as determined from the resolution as a function
of the drift distance for both gases and 3 regions of amplitude.
}
\end{figure}

If the sample is split up in three regions of 
signal amplitude this bias is reduced to about 10\%.
The amplitude ranges are chosen such that the number of events 
in each range is approximately the same.
The fit results are summarized in Table  \ref{tab:transpa} and
 Figure \ref{fig:transpa}. The ratio $R$ is consistent with 0.5 independent
of the amplitude. This interpretation does not account for the 
$\sim$10\% loss of primary electrons over the full drift-distance 
observed in \arco . There is no electron loss in the drift-region for 
P10. No loss of GEM transparency is suggested at our operating voltages
from published analyses \cite{cit:sauli}. 
However, it is clear that the present resolution measurement technique 
effectively uses only about half the statistical power available from the 
number of primary electrons. 

\subsection{X resolution depending on track angle $\phi$}

\begin{figure}[p]
\centerline{\mbox{\epsfxsize=0.9\textwidth \epsffile{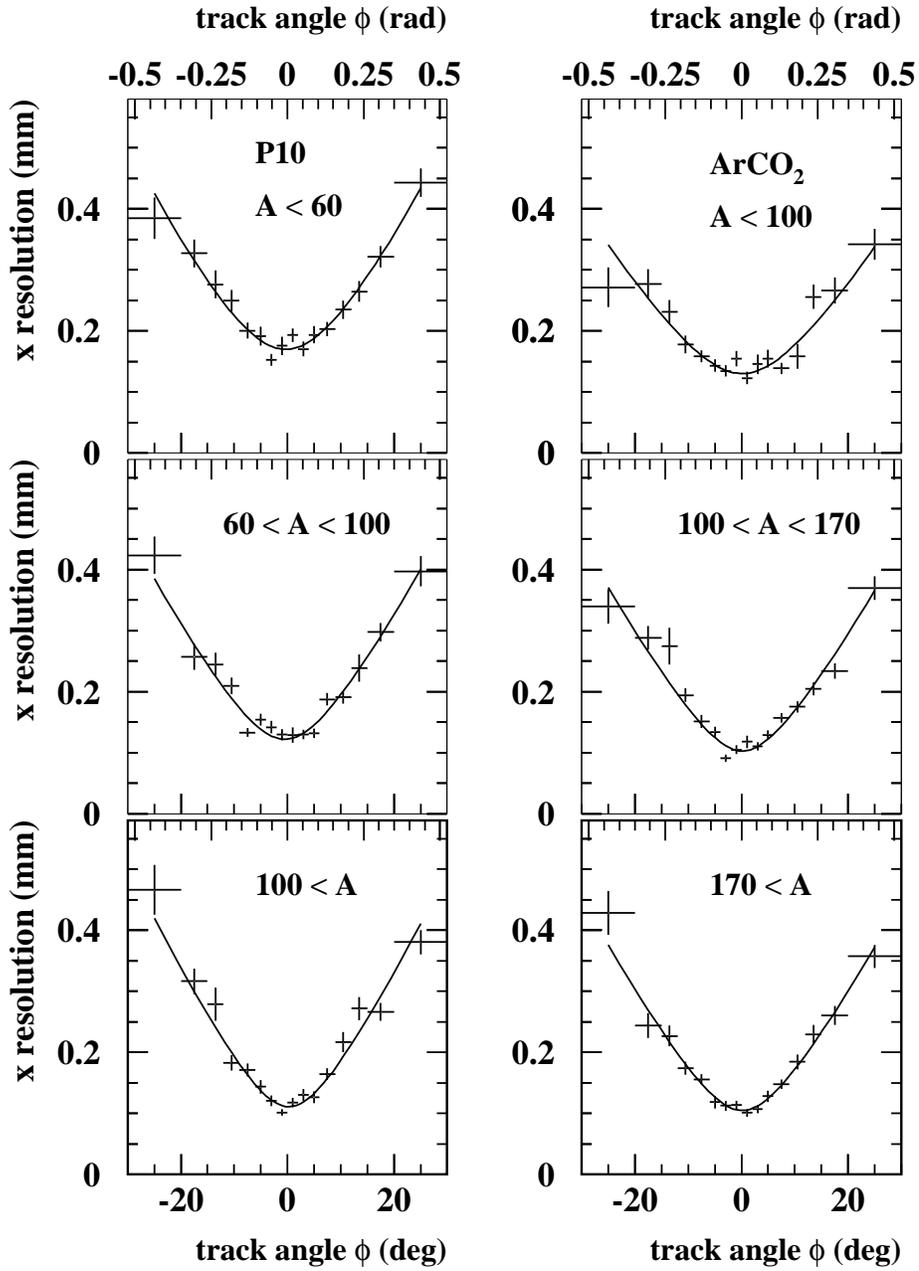}}}
\caption[]{\label{fig:phi_ampl2}
X resolution as a function of track angle for 2 mm wide pads and
drift distance less than 3 cm. For both gases the sample is
split up in three regions of amplitude $A$.
The expected angular dependence is fit to the distributions.
The fit results are given in Table \ref{tab:phifit}.}
\end{figure}

The track angle effect on the
resolution is expected to be $\propto \tan{\phi}$ and depends on
the length of the pad as well as the effective number of 
clusters $N_{\rm cl}^\epsilon$.
Figure~\ref{fig:phi_ampl2} shows the x-resolution as a function of 
track angle for drift distances of less than 3 cm 
for three regions of amplitude.
The following function is fit to the data:
\be
s_{\rm x} = \sqrt{ s^2 + \frac{L^2}{12}\tan^2{(\phi-\varphi)} \; 
/ N_{\rm cl}^\epsilon } \; ,
\ee
where $\varphi$ is an additional
free parameter allowing for a bias in the track angle 
and $s$ includes contributions from diffusion $s_{\rm D}$.
For the fit the number of clusters $N_{\rm cl}$ is taken to be
independent of the amplitude, all variations being assigned to
the exponent $\epsilon$. The fit results for $s$, $\varphi$ and 
$\epsilon$ are given in Table \ref{tab:phifit}.

The offset $\varphi$ is consistent with 0 indicating that there is
no systematic shift.
The number of primary clusters $N_{\rm cl}$ is reduced to the number 
of effective clusters by the exponent $\epsilon$.
We see no dependence of $\epsilon$ on the amplitude.
As expected the basic resolution $s$ improves with amplitude.
Due to the high gain $s$ is mainly determined by the diffusion
$s_{\rm D}$ and not so much by the internal resolution $s_0$.

\begin{table}[b]
\caption[]{\label{tab:phifit}
Fit result for track angle effect.}
\begin{center}
\begin{tabular}{c|r@{$\pm$}lr@{$\pm$}lr@{$\pm$}l}
\hline
\multicolumn{1}{c|}{Amplitude}&
\multicolumn{6}{c}{Fit result}\\
range & 
\multicolumn{2}{c}{$s$ (mm)} & 
\multicolumn{2}{c}{$\varphi$ (deg)} & 
\multicolumn{2}{c}{$\epsilon$} \\
\hline
\multicolumn{7}{l}{P10} \\
$ 0 < A < 60$   & 
0.170&0.005 & --0.3&0.5 &  0.50&0.03 \\
$ 60 < A < 100$  & 
0.122&0.004 & --0.5&0.3 &  0.54&0.02 \\
$ 100 < A < 1000$ & 
0.111&0.004 &   0.3&0.3 &  0.49&0.02 \\
\hline
\multicolumn{7}{l}{\arco} \\
~~$0 < A < 100$~ & 
0.130&0.005 &   0.1&0.5 &  0.64&0.03 \\
$100 < A < 170$~ & 
0.103&0.003 &   0.2&0.3 &  0.56&0.02 \\
$170 < A < 1000$ & 
0.105&0.003 &   0.1&0.3 &  0.55&0.02 \\
\hline
\end{tabular}
\end{center}
\end{table}

\clearpage

\section{Conclusion}

We have investigated the dependence of the spatial resolution of
cosmic-ray tracks in a GEM TPC on various parameters.
We found that the resolution degrades if the pads are too wide with
respect to the track-charge width arriving at the 
readout plane due to insufficient charge sharing between readout pads.
The observed charge width is larger than that expected from transverse 
diffusion between the GEMs and the readout plane.
Therefore 2~mm wide pads are large enough to achieve a resolution
of 100 \mum\ for drift distances of less than 3 cm in the absence of
an axial magnetic field. The dependence of 
the spatial resolution as a function of drift distance allows the
determination of the underlying electron statistics. Our results
show that with the present technique effectively only about half of 
the primary electron statistics is used for the position determination. 
The track angle effect is observed as expected.

\section*{Acknowledgements}
We would like to thank Ron Settles for providing the ALEPH TPC
charge amplifiers that were used in these measurements.
Ernie Neuheimer lent us his expertise in designing, building and
troubleshooting much of our specialized electronics. 
Mechanical engineers Morley O'Neill and Vance Strickland helped with 
the detector design and in improving the clean-room facility where 
the detector was assembled. Philippe Gravelle was always helpful in
providing technical assistance when needed. Much of the work
was done by our CO-OP students Alasdair Rankin, Steven Kennedy,
Roberta Kelly, David Jack and Pascal Elahi
who where involved in construction and commisioning of the detector
as well as in data taking and analysis. 
This research was supported by a project grant from the 
Natural Science and Engineering Research Council of Canada.

\bibliographystyle{Lep2Rep}
\bibliography{pr}

\clearpage

\appendix
\section{Corrections to the $x$ resolution}
\label{sec:rowfit}

When determining the point resolution the typical method involves 
fitting a straight line to all points and determining the standard
deviation of the residuals; this method gives 
a resolution that is too good, since the point for which the 
resolution is to be determined was included in the line fit. 
The alternate method is to fit a straight line without the point 
for which the resolution is to be determined; this gives a resolution 
which is worse than the actual resolution since the line is determined 
from the other points which themselves have an uncertainty.

A detailed analysis reveals that a better estimate of the true 
resolution is given by the geometric mean of the two methods, that is: 
$
\sigma_i^2 = \sigma_{\delta_i^{\rm in}} \cdot \sigma_{\delta_i^{\rm ex}} 
$,
where $\sigma_i$ is the better estimate of the resolution $s_i$ 
for point $i$, $\delta_i^{\rm in}$ and $\delta_i^{\rm ex}$ are 
the measured residuals when the point is included and 
excluded respectively, and $\sigma_{\delta_i^{\rm in}}$ and 
$\sigma_{\delta_i^{\rm ex}}$ are the standard deviations of 
the residual distribution when the point is included 
and excluded from the fit.

Let us assume a track consisting of $N$ measurements with known
values $y_j$, $1\leq j \leq N$. The corresponding measured values
$x_j$ are distributed around the expected mean 
$\langle x_j \rangle = a + b \; y_j$ with the standard deviations $s_j$,
where $a$ and $b$ are the track parameters.
To determine the resolution of one measurement $i$ it is convenient
to choose the coordinate system so that $y_i = 0$.
In this case, the residual is given by $\delta_i = a - x_i$, where
$a$ can be determined from a least square fit to the track by either
including ($a^{\rm in}$) or excluding ($a_i^{\rm ex}$) the 
measurement~$i$. The residual $\delta_i$ will be distributed with a 
standard deviation $\sigma_{\delta_i}$ which is related to $s_i$,
but depends on the coordinates $(x_j,y_j)$ and weights $w_j = 1/s_j^2$ 
of all measurements.


Minimising the $\chi^2$ gives an estimate for $a$:
\ba
a^{\rm in} & = & \frac{\sum_j w_j x_j \cdot \sum_k w_k y_k^2 
                 - \sum_k w_k y_k \cdot \sum_j w_j x_j y_j}
                  {D^{\rm in}} \mbox{ , where} \\ 
\nonumber
D^{\rm in} & = & \sum_j w_j \cdot \sum_j w_j y_j^2 - ( \sum_j w_j y_j )^2 
\ea
and
\ba
a^{\rm ex}_i & = & \frac{\sum_{j \neq i} w_j x_j \cdot 
                     \sum_{k \neq i} w_k y_k^2 
                   - \sum_{k \neq i} w_k y_k \cdot 
                     \sum_{j \neq i} w_j x_j y_j}{D^{\rm ex}_i} \mbox{ , where}\\
\nonumber
D^{\rm ex}_i & = & \sum_{j \neq i} w_j \cdot \sum_{j \neq i} w_j y_j^2 - 
             ( \sum_{j \neq i} w_j y_j )^2 \; .
\ea

And since $y_{(j=i)} = 0$ 
\be
\nonumber
D^{\rm ex}_i = D^{\rm in} - w_i \sum_j w_j y_j^2 \; .
\ee

The residual $\delta_i^{\rm in}$ of point $i$ is:
\ba
\nonumber
\delta_i^{\rm in} & = & a^{\rm in} - x_i \\
\nonumber
 	 & = & \frac{\sum_j w_j x_j  (\sum_k w_k y_k^2 - y_j \sum_k w_k y_k ) 
                   - D^{\rm in} x_i} {D^{\rm in}} \\
 	 & = & \frac{\sum_{j\neq i} w_j x_j  (\sum_k w_k y_k^2 - y_j \sum_k w_k y_k )
                   - D_i^{\rm ex} x_i} {D^{\rm in}} \; .
\ea

Assuming that the $N$ measurements are independent, the variance of the
residual distribution is approximately:
\be
\nonumber
\sigma^2_{\delta_i^{\rm in}} = 
\sum_j \left( \frac{\partial\delta_i^{\rm in}}{\partial x_j} s_j \right)^2  \; .
\ee 
The partial differentiation picks out the $x_j$ terms yielding:
\be
\sigma^2_{\delta_i^{\rm in}} = \frac
 {\sum_{j \neq i} w_j^2 s_j^2
 \left(\sum_k w_k y_k^2 - y_j \sum_k w_k y_k \right)^2 
   - \left( D_i^{\rm ex} s_i\right)^2 }
 {(D^{\rm in})^2} \; .
\ee 
Expanding, rearranging and collecting terms yields: 
%
\be
\sigma^2_{\delta_i^{\rm in}}   =  s_i^2 \frac{D_i^{\rm ex}}{D^{\rm in}} 
\hspace{1cm}
\mbox{and similarly: }
\hspace{1cm}
\sigma^2_{\delta_i^{\rm ex}}  =  s_i^2 \frac{D^{\rm in}}{D_i^{\rm ex}} \; .
\ee

The quantities ${D^{\rm in}}$ and ${D_i^{\rm ex}}$ are fixed for a given 
layout and can be 
calculated to correct the resolution measured, however it is simpler 
to combine the last two expressions:

\be
s_i^2 = \sigma_{\delta_i^{\rm in}} \cdot \sigma_{\delta_i^{\rm ex}} \; .
\ee

With this one can thus get a better estimate of the resolution by 
taking the geometric mean of the resolution as determined with the 
point included in and with the point excluded from the fit without having to 
calculate a correction factor. As expected, our tests show that 
for a large sample the results are identical for the two methods.

\end{document}